\documentclass[12pt,preprint]{aastex}

\slugcomment{Accepted for publication in the Astrophysical Journal}

\shorttitle{Cygnus Loop Nonradiative Shock - \textit{FUSE}}
\shortauthors{Sankrit and Blair}

\begin{document}

\title{\textit{FUSE} Observations of the Cygnus Loop: O~VI Emission
from a Nonradiative Shock \altaffilmark{1}}

\author{Ravi Sankrit \& William P. Blair}
\affil{The Johns Hopkins University}
\affil{Department of Physics and Astronomy,
3400 N. Charles Street,  Baltimore, MD 21218}
\email{ravi@pha.jhu.edu}

\altaffiltext{1}{Based on data obtained for the Guaranteed Time Team by the
NASA-CNES \textit{FUSE} mission operated by the Johns Hopkins University.  Financial
support to U.S. participants has been provided by NASA contract NAS5-32985.}


\begin{abstract}

We present \textit{Far Ultraviolet Spectroscopic Explorer (FUSE)}
observations of a Balmer filament in the northeast region of the Cygnus
Loop supernova remnant.  The data consist of one spectrum obtained
through the $30\arcsec\times30\arcsec$ (LWRS) aperture and three
spectra at adjacent positions obtained through the
$4\arcsec\times20\arcsec$ (MDRS) aperture.  The nonradiative shocks in
the region giving rise to these faint optical filaments produce strong
O~VI~$\lambda\lambda$1032,1038 emission, which is detected in all the
spectra.  The O~VI emission is resolved by \textit{FUSE} into a strong
component centered at 0~km~s$^{-1}$, and weaker components centered at
$\pm~140$~km~s$^{-1}$.  The MDRS spectra allow us to study the
variation of O~VI emission in the post-shock structure.  We find that
the zero velocity emission is associated directly with the Balmer
filament shock, while the high velocity emission comes from a more
uniformly distributed component elsewhere along the line of sight.  We
also find that the shocks producing the emission at
$\pm~140$~km~s$^{-1}$ have velocities between 180~km~s$^{-1}$ and
220~km~s$^{-1}$, if we assume that the ram pressure driving them is the
same as for the zero velocity component shock.  In the context of the
cavity model for the Cygnus Loop, the interaction of the blast wave
with the spherical shell that forms most of the cavity wall can
naturally give rise to the similar red and blue-shifted components that
are observed.

\end{abstract}


\section{Introduction}

Among the most prominent lines in the ultraviolet spectra of several
supernova remnant (SNR) shocks are O~VI~$\lambda\lambda$1032,1038, a
doublet with the ground state as their common lower level.  O~VI is
produced in shocks with velocities of at least 160~km~s$^{-1}$.  For
shocks with velocities up to about 400~km~s$^{-1}$, the strength of the
O~VI lines is very sensitive to the shock velocity.  In addition, the
O~VI line strength also depends on the column density swept up by the
shock.  The regions producing O~VI emission are typically at
temperatures near 300,000~K\@.  Therefore, these lines are the best
tracers of SNR gas that is cooler than the x-ray emitting regions, and
hotter than the regions of bright optical emission.

The optically faint filaments in the northeast region of the Cygnus
Loop SNR produce a large amount of O~VI emission.  In this region, the
optical emission is dominated by Balmer lines, which are collisionally
excited in a narrow zone behind the shock front \citep{che80}.  The
ultraviolet lines, including O~VI~$\lambda\lambda$1032,1038, come from
the hot post-shock gas further downstream where elements are moving to
higher stages of ionization.  The shocked gas has not had time to
recombine and cool radiatively and so these shocks are termed
``nonradiative''.

Strong O~VI emission was detected in a \textit{Hopkins Ultraviolet
Telescope (HUT)} spectrum of a bright nonradiative filament in the
northeast Cygnus Loop by
\citet{lon92}.  They found that the ratio of line strengths
I$_{1038}$/I$_{1032}$ implied significant resonant scattering along the
line of sight, and inferred a pre-shock density between 5 and
12~cm$^{-3}$.  However, they also found that the overall \textit{HUT}
spectrum was best fit by a 180~~km~s$^{-1}$ shock running into material
with a density of 2~cm$^{-3}$.  Other studies of the same filament
based on spectra of the Balmer lines and on \textit{International
Ultraviolet Explorer (IUE)} data also suggest that the preshock density
is $\sim 2$~cm$^{-3}$ \citep{ray83, hes94}.  More recently, analysis of
the distribution of C~IV~$\lambda$1549 and N~V~$\lambda$1240 emission
in a region behind the shock front (seen in a spatially resolved
\textit{Space Telescope Imaging Spectrograph} spectrum) showed that the
pre-shock density is at least 2~cm$^{-3}$ (but probably closer to
4~cm$^{-3}$), and that the shock velocity is $\sim 180$~km~s$^{-1}$
\citep{san00}.

Here, we present \textit{Far Ultraviolet Spectroscopic Explorer (FUSE)}
observations of a portion of this well studied  Balmer filament.  The
only SNR lines detected in the \textit{FUSE} bandpass are
O~VI~$\lambda\lambda$1032,1038.  However, in contrast to the
\textit{HUT} spectrum, the two lines of the doublet are clearly
separated from each other and from the Ly$\beta$ airglow line, and they
are resolved into complex line profiles that vary with spatial position
behind the shock.

\section{Observations}

\textit{FUSE} observations were obtained on 2000 June 6, as part of the
Guaranteed Time Team project on SNRs.  The first observation was
obtained through the low resolution (LWRS) $30\arcsec~\times~30\arcsec$
aperture on the filament (centered at $\alpha_{2000}=20^{\rm{h}}
56^{\rm{m}} 06\fs57$,
$\delta_{2000}=+31\arcdeg\ 56\arcmin\ 05\farcs6$).  Three other
observations were obtained through the medium resolution (MDRS)
20\arcsec~$\times$~4\arcsec\ aperture with the long dimension parallel
to the filament.  The first MDRS position was centered on the same
location as the LWRS observation, and the next two were obtained with
the slit stepped back each time by $\sim$3\arcsec\ perpendicular to its
length.  (Henceforth, we will refer to these positions as P1, P2 and
P3.) Details of the observations are presented in Table \ref{tblobs}.
In Figure~\ref{SLITS}, the aperture locations are shown overlaid on a
WFPC2 H$\alpha$ image of the filament \citep{bla99}.

\textit{FUSE} has four independent channels, LiF1, LiF2, SiC1, and
SiC2, each with two segments, ``A'' and ``B'' (see \cite{moo00} for
details).  Four segments, LiF1A, LiF2B, SiC1A, and SiC2B cover the
wavelength region $\sim$1000 -- 1100\AA, which includes the
O~VI~$\lambda\lambda$1032,1038 doublet.  Of the four segments, LiF1A is
the one with the largest effective area at 1035\AA, and the O~VI lines
are well detected.  Furthermore, since the LiF1A channel is used for
guiding, only the LiF1 apertures can be held accurately in position for
the duration of an observation -- other channels drift with respect to
LiF1 due to thermal motions \citep{sah00}.  Therefore, in this paper we
present only data from the LiF1A segment.

For each observation, we combined the raw data from all individual
exposures and then used CALFUSE pipeline version 1.8.7 to produce
calibrated spectra.  A shift was applied to the flux vectors for each
observed LiF1A spectrum in order to line up geocoronal Ly$\beta$
emission at the appropriate wavelength in the heliocentric frame of
reference.  (The average geocentric to heliocentric velocity is
calculated and included in the headers of the calibrated FITS files.)
The pipeline slightly oversubtracts the background, so we added a
correction to the flux vectors -- this step was done to avoid negative
fluxes in the plots; the measured line strengths are not affected.

\section{Results}

The LWRS spectrum between 1024\AA\ and 1040\AA\ is shown in Figure
\ref{LWRS}.  Each O~VI line has a strong central component, and emission
on both red and blue wings.  The wings are prominent in the
1032\AA\ line but much weaker in the 1038\AA\ line.  The spectral
resolution for emission lines from extended sources is the filled slit
width $\sim0.34$\AA, which corresponds to $\sim100$~km~s$^{-1}$ at
1034\AA\@.  (This is the width of the airglow lines in the spectrum.)
The width of the central component of the O~VI emission is about the
filled slit width.  The integrated line fluxes of the 1032\AA\ and
1038\AA\ lines (including the central component and the wings) are
$7.67\times10^{-13}$~erg~s$^{-1}$~cm$^{-2}$ and
$3.66\times10^{-14}$~erg~s$^{-1}$~cm$^{-2}$, respectively.

The MDRS spectra between 1024\AA\ and 1040\AA\ are shown in Figure
\ref{MDRS}.  The filled slit spectral resolution with this aperture is
$\sim0.045$\AA, which corresponds to $\sim13$~km~s$^{-1}$ at
1034\AA\@.  The O~VI lines are broader than the filled slit width
(compare the O~VI lines with the airglow lines).  The two components on
the wings of the 1032\AA\ line are clearly seen in each of the
spectra.  The O~VI line strengths decrease from P1 to P3.  Also, in the
P3 spectrum, the lines are broader and the central component of
O~VI~$\lambda$1032 is double peaked.  The wing components are about
equally strong at all three positions.  The O~VI line fluxes in the
MDRS spectra, along with the fluxes in the LWRS spectrum are given in
Table \ref{tblflux}.  The fluxes were obtained by simple trapezoidal
integration over the line profiles.  The error in each measurement is
dominated by the absolute flux calibration, which is accurate to 10\%
\citep{sah00}.

The O~VI line fluxes at the three MDRS positions are plotted against
velocity in Figure \ref{VEL}.  The top panel shows O~VI~$\lambda$1032
and the bottom panel shows O~VI~$\lambda$1038.  The central components
of the 1032\AA\ line and the 1038\AA\ line are centered at
0~km~s$^{-1}$, and their FWHMs are $\sim100$~km~s$^{-1}$.  The
components on the 1032\AA\ line wings are centered at approximately
$\pm140$~km~s$^{-1}$.

The intrinsic line profiles of O~VI~$\lambda$1032 and
O~VI~$\lambda$1038 are expected to be identical, except that the
1032\AA\ line is twice as strong as the 1038\AA\ line in the optically
thin limit.  The features on the wings of the 1032\AA\ line in the \textit{FUSE}
spectra are not likely to be spurious since we see them in the LiF2B
spectra as well.  (The LiF2B spectra are not shown here.)  Their absence
in the 1038\AA\ emission can be attributed to absorption by molecular
hydrogen along the line of sight between us and the Cygnus Loop.  The
Lyman band transitions of H$_{2}$, R(1)$_{5-0}$ and P(1)$_{5-0}$, have
rest wavelengths of 1037.146\AA\ and 1038.156\AA, respectively
\citep{mor76}.  Relative to the central O~VI wavelength (1037.617\AA)
these are at $-136$~km~s$^{-1}$ and $+155$~km~s$^{-1}$, respectively.
If H$_2$ is present along the line of sight, then we expect these
strong transitions to affect the spectrum strongly around the O~VI
1038\AA\ line.

To check this quantitatively, we used a \textit{FUSE} spectral simulation
routine, FSIM, to examine the effects of H$_{2}$ absorption.  O~VI
emission components at zero velocity and at $\pm140$~km~s$^{-1}$ were
included.  The peak fluxes on the 1032\AA\ wings were chosen to be
$4\times10^{-14}$~erg~s$^{-1}$~cm$^{-2}$~\AA$^{-1}$; the peak fluxes of
the 1038\AA\ line were chosen to be half this value (valid for
optically thin line emission).  We find that $10^{16}$~cm$^{-2}$ of
H$_{2}$ at a temperature of 250~K would be sufficient to absorb the
O~VI~$\lambda$1038 wings to the point that the emission would be
undetectable.  Although we do not attempt to set any limits, we
conclude that there is sufficient molecular hydrogen towards the Cygnus
Loop that results in the O~VI~$\lambda$1038 wings being absorbed.
(O~VI line profiles showing the presence of overlying H$_2$ absorption
are also seen in \textit{FUSE} spectra of other regions in the Cygnus Loop
(Blair et al.~in preparation).)

\section{Discussion}

The set of MDRS spectra were obtained in order to map the spatial
distribution of the O~VI emission perpendicular to the shock front.  We
find that the central component of O~VI emission is well correlated
with the H$\alpha$ emission from the filament.  The O~VI flux is
highest at P1, where the filament is most edge-on and the H$\alpha$ is
brightest (Figure \ref{SLITS}), and is lower at positions P2 and P3,
which lie behind the leading edge of the filament.  Furthermore, the
O~VI and H$\alpha$ emission are both centered at 0~km~s$^{-1}$ (Figure
\ref{VEL}, and \citet{hes94}).  The high velocity components, on the
other hand, do not vary much among the three MDRS positions (Table
\ref{tblflux}) and we infer that they come from regions that are more
uniformly distributed (in the plane of the sky) than the gas behind the
optical filament.

The O~VI surface brightnesses measured in the \textit{FUSE} spectra and in the
HUT spectrum are listed in Table \ref{tblsb}.  The HUT observation was
obtained with the $9\farcs4\times116\arcsec$ placed with its length
along the Balmer filament (see Figure 1 of \citet{lon92}).  At the
distance of the Cygnus Loop -- 440~pc \citep{bla99}, the aperture
length corresponds to $\sim0.25$~pc.  The O~VI surface brightness of
the region observed by HUT is as high as it is at P1.  Therefore, we
infer that strong O~VI emission is more or less uniformly distributed
along the length of the filament.  Spatial variations in the O~VI flux
are much larger in the direction perpendicular to the shock front.

Approximately 1/3 the area of the LWRS aperture covers a region that is
ahead of the H$\alpha$ filament (Figure \ref{SLITS}).  Therefore it is
not surprising that the O~VI surface brightness in the LWRS spectrum is
even lower than the surface brightness in the MDRS P3 spectrum.  The
average surface brightnesses of the three MDRS apertures are
$13.5\times10^{-16}$~erg~s$^{-1}$~cm$^{-2}$~arcsec$^{-2}$ for
O~VI~$\lambda$1032 and
$6.9\times10^{-16}$~erg~s$^{-1}$~cm$^{-2}$~arcsec$^{-2}$ for
O~VI~$\lambda$1038.  If we assume that 2/3 of the LWRS aperture is
filled with O~VI, we can calculate corrected LWRS surface brightnesses
by multiplying the values in Table \ref{tblsb} by 3/2.  The corrected
surface brightnesses are
$12.8\times10^{-16}$~erg~s$^{-1}$~cm$^{-2}$~arcsec$^{-2}$ and
$6.2\times10^{-16}$~erg~s$^{-1}$~cm$^{-2}$~arcsec$^{-2}$ for the
1032\AA\ and 1038\AA\ lines, respectively.  These numbers are much
closer to the average MDRS brightnesses.  This analysis suggests
that there is very little O~VI emitted in the region just ahead
of the filament.

By comparing various spectra taken by \textit{FUSE} and HUT, we have inferred
some of the properties of the distribution of O~VI emission in the
vicinity of the Balmer line filament.  Now, we consider the properties
of the emission components in the MDRS spectra and what can be learned
from them about the shock and cloud properties.


\subsection{The Zero Velocity Component}

The distribution of O~VI emission seen in the three MDRS spectra is
affected by the history and the geometry of the shock
front.  According to our current understanding of the Cygnus Loop,
based on the the cavity model \citep{hes94,lev98}, the shock now giving
rise to the Balmer filaments encountered the cavity ``wall'' several
hundred years ago.  Although the radial density profile of this
``wall'' is not known, it is likely that the pre-shock density was
lower and the shock velocity higher at earlier times.  The contribution
to the O~VI flux from these earlier times is uncertain, but probably
not very high: a fast shock will ionize oxygen past O$^{5+}$ rapidly,
and the lower pre-shock density implies a lower O~VI yield.  Hence it is
likely that the differences in the zero velocity component of O~VI
emission among the three MDRS spectra are due mainly to the shock
geometry.  In this paper, we will not consider the shock history
further.

The shock front is a rippled sheet with a peak to peak amplitude of
$\sim$10\arcsec, measured in the WFPC2 H$\alpha$ image of the filament
\citep{bla99}.  On larger scales, the shock front is curved with the
convex face outward \citep{lev98}.  It may be assumed that the filament
similarly curves along the line of sight.  The O~VI flux is highest at
P1 because it is closest to the edge -- i.e. it is
``limb-brightened''.  The flux at position P3 is lower because the path
length through the emitting gas is shorter than at P1.  Because the
path length is shorter, the effects of resonance scattering are smaller
and the I$_{\lambda1038}$/I$_{\lambda1032}$ is closer to 0.5 at P3 than
at P1 (Table \ref{tblflux}).  The O~VI lines are slightly broader in
the P3 spectrum than in the P1 and P2 spectra, and the 1032\AA\ line
shows a double peak.  This is due to the curvature of the filament --
much of the emission at P3 comes from portions of the shock front that
have a small radial velocity component.  

We have accounted for the spatial variation of the O~VI emission seen
in the MDRS spectra in a qualitative way using geometric arguments.
However, the spatial resolution and coverage of the observations are
insufficient to distinguish between different values of the shock
velocity and pre-shock density.  Therefore, in the following
discussion, we use the average O~VI surface brightness of the three
MDRS positions.

The total flux of the O~VI$\lambda$1032 zero velocity component in the
MDRS spectra (P1,P2 and P3) is
$2.72\times10^{-13}$~erg~s$^{-1}$~cm$^{-2}$ (Table \ref{tblflux}),
which corresponds to a surface brightness of
$1.13\times10^{-15}$~erg~s$^{-1}$~cm$^{-2}$~arcsec$^{-2}$.  In order to
compare the observed flux with shock model predictions, two corrections
have to be made.  First, the interstellar extinction has to be
accounted for.  We use the extinction curve presented by \citet{fit99}
and find, for color excess E$_{B-V}$=0.08 to the Cygnus Loop
\citep{fes82} and $R_V = 3.1$ (the standard ISM value), that the
correction factor is 2.8 at 1032\AA\@.  Second, since we are viewing
the shock close to edge-on, the resonance scattering of the O~VI lines
can significantly attenuate their intensities.  If resonant scattering
effects were very small (i.e.~the optical depth in the lines was
negligible) then the ratio I$_{\lambda1038}$/I$_{\lambda1032}$ would be
0.5.  The observed ratio is about 0.60, which implies that the optical
depth in the 1032\AA\ line is $\sim0.8$.  Following the procedure
outlined in \citet{lon92}, we find that the attenuation factor for the
line is $\sim1.5$.  The corrected surface brightness of the central
component of O~VI$\lambda$1032 (obtained by multiplying the measured
surface brightness by these correction factors derived above) is
$4.7\times10^{-15}$~erg~s$^{-1}$~cm$^{-2}$~arcsec$^{-2}$.

We choose parameters based on earlier studies of the filament
\citep{ray83,lon92,hes94,san00} and use these in shock models to see
how the predictions compare with the observation.  Three models were
calculated, with shock velocities of 170, 180 and 190~km~s$^{-1}$.
Diffuse ISM abundances, where O:H = 8.70:12.00 on a logarithmic scale
\citep{cow86}, and a pre-shock hydrogen number density of 3~cm$^{-3}$
were used in all the models.  The models were calculated using an
updated version of the code described by \citet{ray79}.  The models
predict the face-on O~VI intensity as a function of the swept up column
density.  

In order to compare the predicted intensity with the observed
intensity it is necessary to take into account the viewing aspect
ratio, i.e.~the ratio of actual shock surface area lying within the
slit to the projected area.  The observed region is
10\arcsec$\times$20\arcsec (the MDRS positions overlap - see Figure
\ref{SLITS}).  If the extent of the shock front along the line of sight
is $l$ arcseconds, then since the MDRS apertures are placed parallel to
the filament, the aspect ratio is $\sim~l/10$.  Figure 1 of
\citet{hes94} shows that the filament is about 300\arcsec\ long.  If we
assume that the extent of the filament along the line of sight is
comparable to its extent on the sky (there is no reason to believe
otherwise), then the aspect ratio is about 30.  Now, the filament we
are studying is the brightest Balmer filament in the northeast Cygnus
Loop.  This may be a selection effect based on a longer path length
along the line of sight so in the following discussion we also
consider the possibility of a higher aspect ratio as well.

In Figure \ref{MODS} the O~VI intensities predicted by the shock models
are plotted as functions of swept up hydrogen column density.  The
horizontal lines show the observed intensities (corrected for
interstellar extinction and resonance scattering as described above)
for aspect ratios of 30 and 60, which may be considered as approximate
bounds on the true value.  Note that these models are all calculated
for pre-shock density, n$_{0}=3$~cm$^{-3}$; for a given shock velocity
and swept up column density, the O~VI intensity scales linearly with
n$_{0}$.  Analyses of other ultraviolet spectra \citep{lon92, san00}
have shown that the swept up column density lies in the range
$\sim0.8$--$2\times10^{17}$~cm$^{-2}$.  From the plot we see that the
models predict the O~VI emission correctly.  The important conclusion
of the preceding analysis is that the observed O~VI flux can be
accounted for by nonradiative shock emission alone -- it is not necessary to
invoke other mechanisms such as cloud evaporation, or ``coronal''
emission from cooling gas ionized at an earlier time.

The 180~km~s$^{-1}$ shock model predicts a total O~VI line intensity
$\sim16$ times the H$\alpha$ intensity for a swept up column of
$1\times10^{17}$~cm$^{-2}$.  Although only the O~VI doublet lines lie
in the \textit{FUSE} bandpass, other strong UV lines have been detected
in spectra of this filament.  Predicted intensities of these UV lines
from similar shock models are given by \citet{lon92}.  Several lines of
He~I, He~II, O~IV and O~V at wavelengths below 912\AA\ are predicted to
be comparable to O~VI in intensity.  The strongest of these are
He~II~$\lambda$304 (a few times as strong as O~VI), and
O~V~$\lambda$630 (about 1.5 times as strong as O~VI).  These lines
cannot be observed because of interstellar absorption, but they can be
important for photoionization of the pre-shock gas \citep{ham88}.

A few lines in the optical range are expected to be present at the
level of a few percent the brightness of H$\alpha$, in particular
He~I~$\lambda$6678, He~II~$\lambda$4686 \citep{har99} and
[O~III]~$\lambda$5007.  The brightest predicted line,
[Ne~V]~$\lambda$3425 at about 10\% the H$\alpha$ intensity, was
detected by \cite{ray83}.  Infrared emission lines are predicted to be
very faint, the brightest being [O~IV]~25.9$\mu$m and [Ne~VI]~7.6$\mu$m
lines at about 1\% the intensity of H$\alpha$.


\subsection{The Wing Components}

The total flux of the high velocity O~VI at the three MDRS positions is
$5.3\times10^{-14}$ erg s$^{-1}$ cm$^{-2}$ (Table \ref{tblflux}).  The
correction for interstellar extinction is exactly the same as for the
main component (a factor of 2.8).  As we discuss below, resonance
scattering effects may be neglected.  So, the corrected surface
brightness for the wing components (red and blue total) is
$6.2\times10^{-16}$~erg~s$^{-1}$~cm$^{-2}$~arcsec$^{-2}$.  It is seen
from Figure \ref{VEL} that flux in the blue wing is approximately the
same as the flux in the red wing.  To check this quantitatively, we fit
gaussians to the wing profiles, and also evaluated the integrals under
the curve, between suitably chosen points.  Both methods showed that
the difference in flux between the two wings is $\lesssim$10\%.
For the following discussion, we assume that the two components are
equally bright and consider only the blue wing -- the arguments used
and results derived apply equally well to both components.

We start by making the assumption that the $-140$~km~s$^{-1}$ emission
is due to a steady shock traveling towards us, and that we are looking
through just a single shock structure.  (The line optical depth will be
small, and attenuation of the flux due to resonance scattering can be
neglected.) The angle between the shock normal and the line of sight
($\theta$), the observed and face-on surface brightnesses, and the
observed velocity and shock velocity are related by
\begin{equation}
cos\theta = \frac{v_{obs}}{v_s} = \frac{I_0}{I_{obs}}
\end{equation}
and the latter relation may be written as
\begin{equation}
I_0 v_s = I_{obs} v_{obs}.
\end{equation}
Thus, there are a family of solutions for the shock velocity and
face-on surface brightness that will match the observations.

We next assume that the pressure driving this shock is equal to the
pressure driving the shock responsible for the H$\alpha$ filament and
the zero velocity O~VI emission.  This is a reasonable assumption since
ROSAT x-ray maps show no evidence for large pressure variations in the
region \citep{lev97}.  Then the isobaric relation, $n_0
v_s^2~=~constant$, holds.  The constant, based on the parameters used
in the model described above (\S4.1), is
$3~\times~180^2$=97200~(cm$^{-3}$)(km~s$^{-1}$).  The isobaric relation
can be used to specify the pre-shock density for every assumed value of
shock velocity.

As discussed in \S4.1, for a given shock velocity and pre-shock
density, the O~VI intensity
depends on the swept up column density.  A third constraint on
the shock parameters is obtained from the relation
\begin{equation}
n_0 v_s t_{shock} = N_{\rm{H}}.
\end{equation}
Here $t_{shock}$ is the time since the shock interaction started (the
age of the shock), and $N_{\rm{H}}$ is the swept up column density.

We ran a sequence of models with shock velocities 160 --
300~km~s$^{-1}$.  The pre-shock densities were calculated using the
isobaric condition.  Equation 2 was used to find the required O~VI
surface brightness for each shock velocity.  For each model this was
achieved at some value of the swept up column density.  (The exception
is the 160~km~s$^{-1}$ shock, which does not produce enough O~VI
emission to match the observations.) The age of the shock is then found
from equation 3.  In Table \ref{tblmod} we list the parameters used in
the models, the required O~VI~1032\AA\ surface brightnesses, and the
resulting swept up column density and shock age.  The swept up column
density required to produce the observed O~VI increases with shock
velocity.  As a consequence of the isobaric condition and equation 3,
$t_{shock}\propto N_{\rm{H}}v_s$, so the shock age also increases with
shock velocity.  There is a sudden jump in the required $N_{\rm{H}}$,
and hence the shock age between shock velocities 200 and
220~km~s$^{-1}$.  This is because in faster shocks, the oxygen rapidly
ionizes to O$^{6+}$ and beyond and the required O~VI intensity is
produced only after the gas recombines to O$^{5+}$.

The age of the Cygnus Loop is $\sim9,000\times(D_{pc}/440)$~yr, where
$D_{pc}$ is the distance to the remnant \citep{lev98}.  Since the shock
age cannot exceed the age of the remnant, shock velocities
$\gtrsim260$~km~s$^{-1}$ are ruled out (Table \ref{tblmod}).  However,
more stringent limits can be placed on the shock velocity because
detailed studies of the northeast Cygnus Loop \citep{hes94} show that
the interaction started $\sim1000$~yr ago.  Therefore, the allowed
range of shock velocities is 180--220~km~s$^{-1}$ (Table
\ref{tblmod}).  The angles between the line of sight and the shock
normal for the lower and upper velocity limits are 39\arcdeg\ and
50\arcdeg, respectively (equation 1).

We note that the exact upper limit on the shock velocity derived above
depends on the value of the constant in the isobaric relation.  It also
depends upon the observed O~VI surface brightness, for which the
error is dominated by the uncertainty in our knowledge of the
reddening correction at 1032\AA\@.  However, the expected range of
values for these parameters is sufficiently narrow that though the
limit may be revised upward by some tens of km~s$^{-1}$, the basic
picture for the production of O~VI emission would not change.

Our analysis has shown that either high velocity component of O~VI can
be produced by a shock with properties similar to the one producing the
Balmer line emission.  The similarity between the blue and red wing
components (equal flux, symmetric about zero in velocity space) can be
understood in the context of the cavity model for the Cygnus Loop.  In
this model \citep{hes94,lev98}, the blast wave is interacting with the
walls of the cavity cleared by the progenitor star.  According to the
picture presented by \citet{lev98}, the cavity wall consists of a
neutral atomic shell covering most of the surface, and several large
cloud boundaries filling in the rest.  The shell was formed by
recombination at the edges of the pre-supernova H~II region.  In such a
model we expect to see symmetrical red- and blue-shifted shocks along
lines of sight that do not intersect the large clouds.  Thus, in the
high velocity wings observed with \textit{FUSE} we are seeing the shock
wave interacting with the neutral shell.  Because the shell is
spherically symmetric, our line of sight will pass through shock fronts
inclined at equal angles toward us and away from us.

\section{Concluding Remarks}

We have detected strong O~VI emission from a Balmer filament in the
Cygnus Loop.  The \textit{FUSE} data presented here have the spectral
resolution not only to separate the two O~VI lines from each other and
from the Ly$\beta$ airglow, but also to separate kinematically
different emission components along the line of sight.  The O~VI flux
distribution, sampled at a spatial resolution of $\sim$4\arcsec, is
correlated with the H$\alpha$ emission.  A nonradiative shock with
properties derived from other ultraviolet observations can produce the
observed O~VI\@.  We have also resolved the O~VI emission in the form
of wings on the main component.  These high velocity emission
components are centered symmetrically in velocity space, at
$\pm140$~km~s$^{-1}$, and have equal flux.  They probably arise
in shocks driven into the spherically symmetric neutral shell,
which, in some cavity models, has a large covering factor around
the surface of the remnant.

O~VI~$\lambda\lambda$1032,1038 emission is an important channel through
which the Cygnus Loop SNR loses energy.  Measurements of the global
O~VI emission have been obtained using \textit{Voyager 1} and
\textit{2} \citep{bla91, van93}, and a rocket-borne experiment
\citep{ras92}.  These studies show that the O~VI luminosity is about
the same as the luminosity in the 0.1--4.0~keV x-ray band, the latter
based on \textit{Einstein} observations \citep{ku84}.  The contribution
of nonradiative shocks to the total O~VI emission may be estimated
based on our results.  The face-on surface brightness of the shock
causing the observed wing emission is
$\sim3\times10^{-16}$~erg~s$^{-1}$~cm$^{-2}$~arcsec$^{-2}$ (total of
1032\AA\ and 1038\AA\ lines).  Assuming a covering factor of 50\% for
such shocks, and distance to the Cygnus Loop of 440~pc, the derived
O~VI luminosity is $\sim1.2\times10^{36}$~erg~s$^{-1}$.  While
radiative shocks are known to contribute to the total O~VI emission
\citep{bla91b,dan00}, nonradiative shocks could account for most of the
total O~VI emitted by the Cygnus Loop.  The study of O~VI emission is
clearly important for unraveling the details of the interaction between
the Cygnus Loop and the surrounding material, and by extension for
understanding the nature of other middle-aged cavity SNRs.

\acknowledgements

We thank the people who worked on the development of \textit{FUSE} and
those operating the satellite, and we thank Bryce Roberts for help with
tracking down the accurate slit position angles from the Mission
Planning files.  We thank John Raymond for useful discussions, and for
giving us a copy of his shock code to use for model calculations.  RS
thanks Nancy Levenson for useful discussions.  We acknowledge financial
support provided by NASA contract NAS5-32985 to the Johns Hopkins
University.


\clearpage
\figcaption{Aperture locations overlaid on a WFPC2 H$\alpha$ image of
the Balmer filament in the Cygnus Loop.  The large box (dashed line)
represents the 30\arcsec$\times$30\arcsec\ LWRS aperture and the
smaller boxes represent the 4\arcsec$\times$20\arcsec\ MDRS aperture.
The shock direction is shown on the image.
             \label{SLITS} }
\figcaption{LWRS spectrum of the filament showing the region
around the O~VI~$\lambda\lambda$1032,1038 lines.
The line at about 1025.8\AA\ is Ly$\beta$, and the other airglow
lines are from O~I\@.  The feature on the blue wing of the
Ly$\beta$ is a detector artifact.
             \label{LWRS} }
\figcaption{MDRS spectra showing the region around the O~VI lines.
Position 1 refers to the aperture location that is furthest ahead, on
the optical filament (Figure \protect\ref{SLITS}).  Positions 2 and 3
are stepped back perpendicular to the shock normal.  All spectra
have been binned by 4 pixels.  Airglow lines are marked on the
Position 1 spectrum.
             \label{MDRS} }
\figcaption{O~VI line fluxes plotted as a function of velocity for the
three spectra.  The top panel shows the 1032\AA\ line and the bottom
panel shows the 1038\AA\ line.  The scales are the same on both plots
so they can be compared easily.  The units of F$_{\lambda}$ are
erg~s$^{-1}$~cm$^{-2}$~\AA$^{-1}$.  In the bottom panel, the positions
of a pair of H$_2$ Lyman band transitions are shown.  These H$_2$
transitions are responsible for absorbing the O~VI~$\lambda$1038 wing
emission.
             \label{VEL} }
\figcaption{The cumulative face-on O~VI $\lambda$1032 intensities
predicted by shock models plotted versus swept up hydrogen column
density.  Predictions from three models, with shock velocities of 170,
180 and 190~km~s$^{-1}$ are shown.  All models use a pre-shock hydrogen
number density of 3~cm$^{-3}$.  The intensity for a given shock
velocity and swept up column scales linearly with the pre-shock
density.  The horizontal lines show the observed intensity divided by
30 and 60, to account for the viewing aspect ratio of the shock front.
Intensity units are: $10^{-16}$~erg~s$^{-1}$~cm$^{-2}$~arcsec$^{-2}$.
             \label{MODS} }

\clearpage
\begin{deluxetable}{ccccc}

\tablecaption{List of Observations
                              \label{tblobs}}
\tablewidth{0pt}
\tablehead{
  \colhead{Obs. ID} & \colhead{$\alpha_{J2000}$} &
  \colhead{$\delta_{J2000}$} & \colhead{t$_{exp}$ (s)} & \colhead{Aperture}
}
\startdata
P1140401  &  $20^{\rm{h}}56^{\rm{m}}06\fs57$ & +31\arcdeg\ 56\arcmin\ 05\farcs6
          &  10234  & LWRS  \\
P1140402  &  $20^{\rm{h}}56^{\rm{m}}06\fs57$ & +31\arcdeg\ 56\arcmin\ 05\farcs6
          &  9559 & MDRS (P1)  \\
P1140501  &  $20^{\rm{h}}56^{\rm{m}}06\fs40$ & +31\arcdeg\ 56\arcmin\ 03\farcs5
          &  9247 & MDRS (P2)  \\
P1140601  &  $20^{\rm{h}}56^{\rm{m}}06\fs23$ & +31\arcdeg\ 56\arcmin\ 01\farcs4
          &  9767 & MDRS (P3)  \\
\enddata

\tablecomments{The slit position angle was 308\fdg5 (E of N) for all
observations.}
\end{deluxetable}

\clearpage
\begin{deluxetable}{lcccc}

\tablecaption{Observed Fluxes and Flux Ratios 
                                   \label{tblflux}}
\tablewidth{0pt}
\tablehead{
  \colhead{Component} & \colhead{MDRS P1} & 
  \colhead{MDRS P2} & \colhead{MDRS P3} & \colhead{LWRS}
}
\startdata
1032\AA\ total   &  1.37  &  1.12  &  0.76  &  7.67   \\
1032\AA\ center  &  1.22  &  0.95  &  0.55  & \nodata \\
1038\AA\ \tablenotemark{a}  &  0.79  &  0.54  &  0.31  &  3.66 \\
1032\AA\ wings\tablenotemark{b} & 0.15 & 0.17 & 0.21 & \nodata  \\
\\
I$_{1038}$/I$_{1032(center)}$  & 0.65 & 0.57 & 0.56 & 0.48\tablenotemark{c} \\
\enddata

\tablecomments{Flux units: 10$^{-13}$ erg s$^{-1}$ cm$^{-2}$.}
\tablenotetext{a}{The central component dominates, as the
wings are absorbed by interstellar H$_{2}$ lines.}
\tablenotetext{b}{This is simply the difference between the
total and center fluxes given in the first two rows.}
\tablenotetext{c}{The components have not been deconvolved in the
LWRS spectrum so the ratio I$_{1038}$/I$_{1032(total)}$ is tabulated.}

\end{deluxetable}

\clearpage
\begin{deluxetable}{lccccc}

\tablecaption{Observed Surface Brightnesses
                                   \label{tblsb}}
\tablewidth{0pt}
\tablehead{
  \colhead{Component} & \colhead{MDRS P1} & \colhead{MDRS P2} &
  \colhead{MDRS P3} & \colhead{LWRS} & \colhead{HUT\tablenotemark{a}}
}
\startdata
1032\AA\ total   & 17.1 & 14.0 &  9.5 &  8.5 & 16.8 \\
1038\AA\         &  9.9 &  6.8 &  3.9 &  4.1 & 10.1\tablenotemark{b} \\
\enddata

\tablecomments{Surface Brightness 
units: 10$^{-16}$ erg s$^{-1}$ cm$^{-2}$ arcsec$^{-2}$.}
\tablenotetext{a}{Data from Long et al.~(1992).  The HUT aperture
was $9\farcs4 \times 116\arcsec$.}
\tablenotetext{b}{Long et al.~(1992) reported the total flux in both
lines of the doublet.  The best fit to their spectrum yielded
a ratio I$_{1038}$/I$_{1032}$=0.6, which we use here.}

\end{deluxetable}

\clearpage
\begin{deluxetable}{ccccr}

\tablecaption{Model Parameters and Results
                                   \label{tblmod}}
\tablewidth{0pt}
\tablehead{
  \colhead{$v_s$} & \colhead{$n_0$} & \colhead{$I_0$\tablenotemark{a}} &
  \colhead{$N_{\rm{H}}$} & \colhead{$t_{shock}$} \\
  \colhead{(km s$^{-1}$)} & \colhead{(cm$^{-3}$)} &
  \colhead{} &
  \colhead{(cm$^{-2}$)} & \colhead{(years)}
}
\startdata
160 & 3.80 & 2.71 & \nodata             & \nodata  \\
180 & 3.00 & 2.41 & $1.86\times10^{17}$ &   110    \\
200 & 2.43 & 2.17 & $2.21\times10^{17}$ &   140    \\
220 & 2.01 & 1.97 & $1.52\times10^{18}$ &  1090    \\
240 & 1.69 & 1.81 & $5.65\times10^{18}$ &  4420    \\
260 & 1.44 & 1.67 & $1.26\times10^{19}$ & 10650    \\
280 & 1.24 & 1.55 & $2.18\times10^{19}$ & 19890    \\
300 & 1.08 & 1.45 & $3.29\times10^{19}$ & 32140    \\
\enddata

\tablenotetext{a}{Units are $10^{-16}$ erg s$^{-1}$ cm$^{-2}$ arcsec$^{-2}$.}

\end{deluxetable}

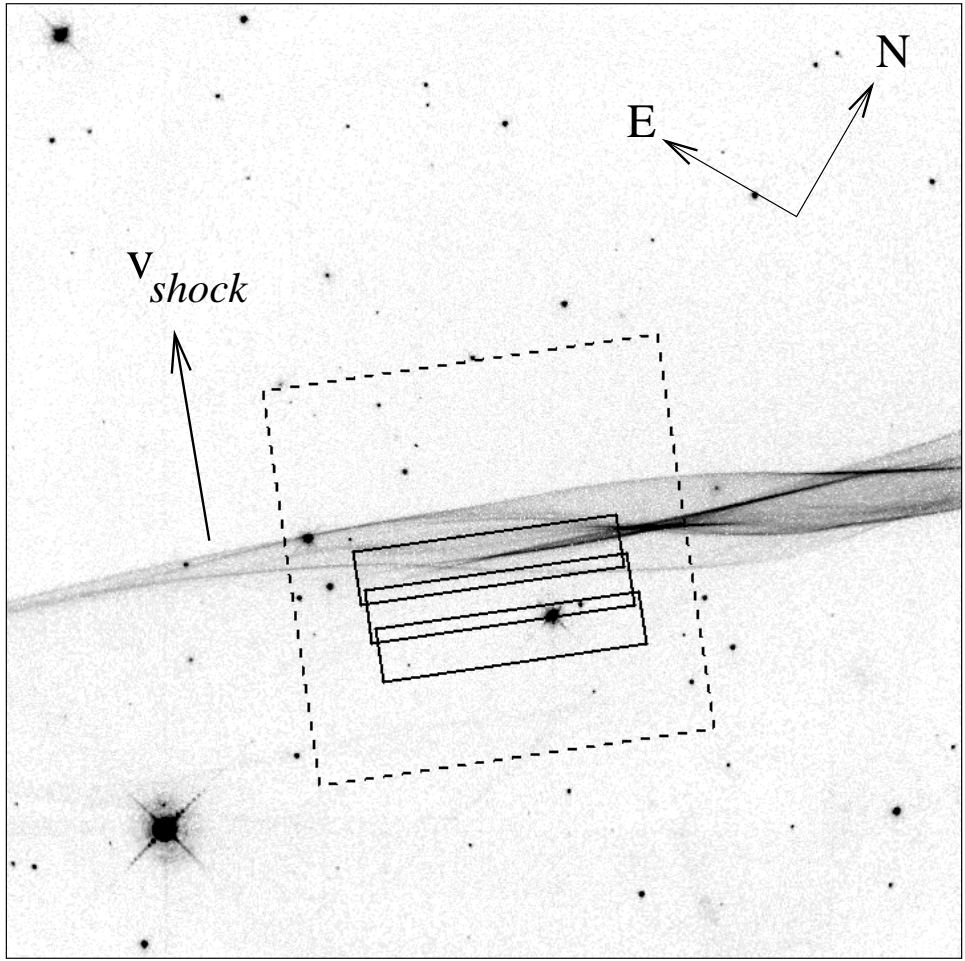
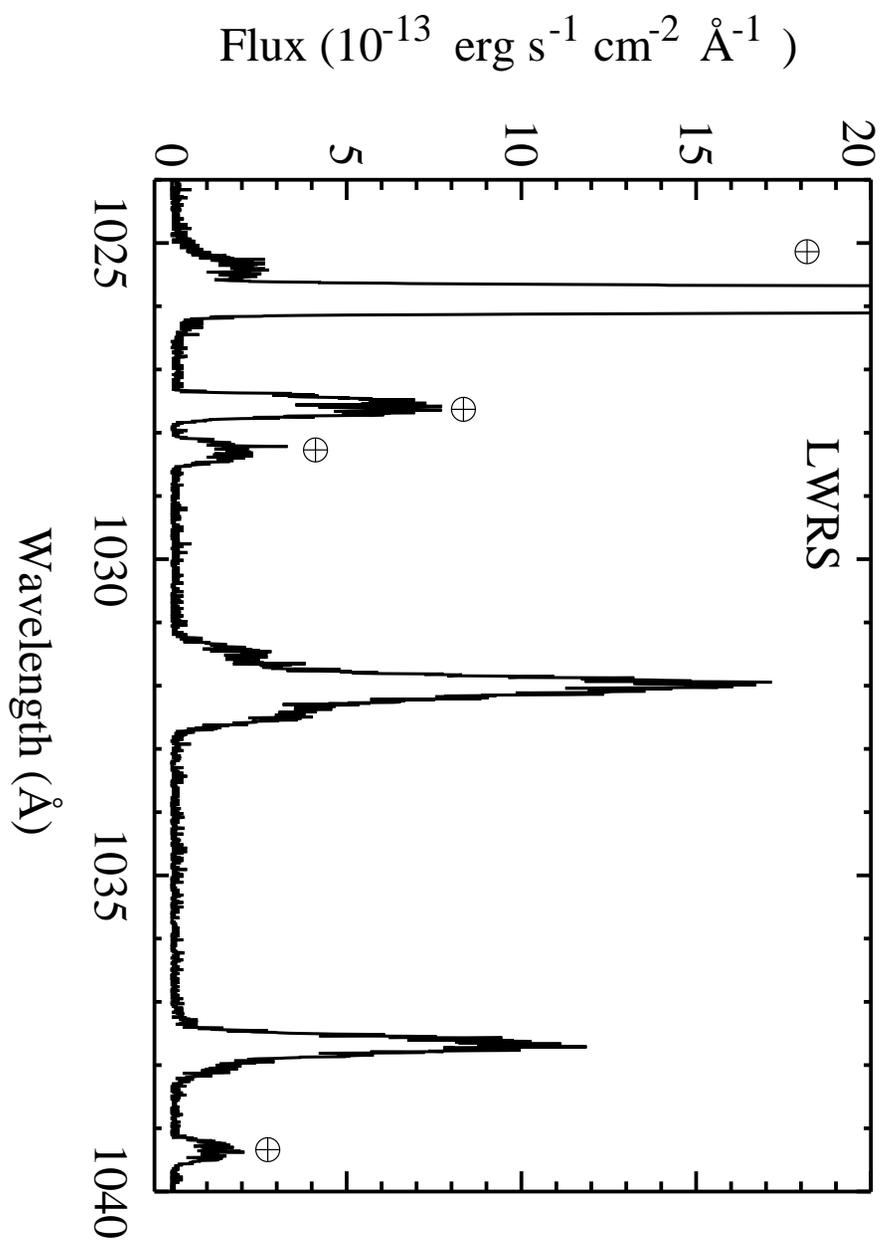
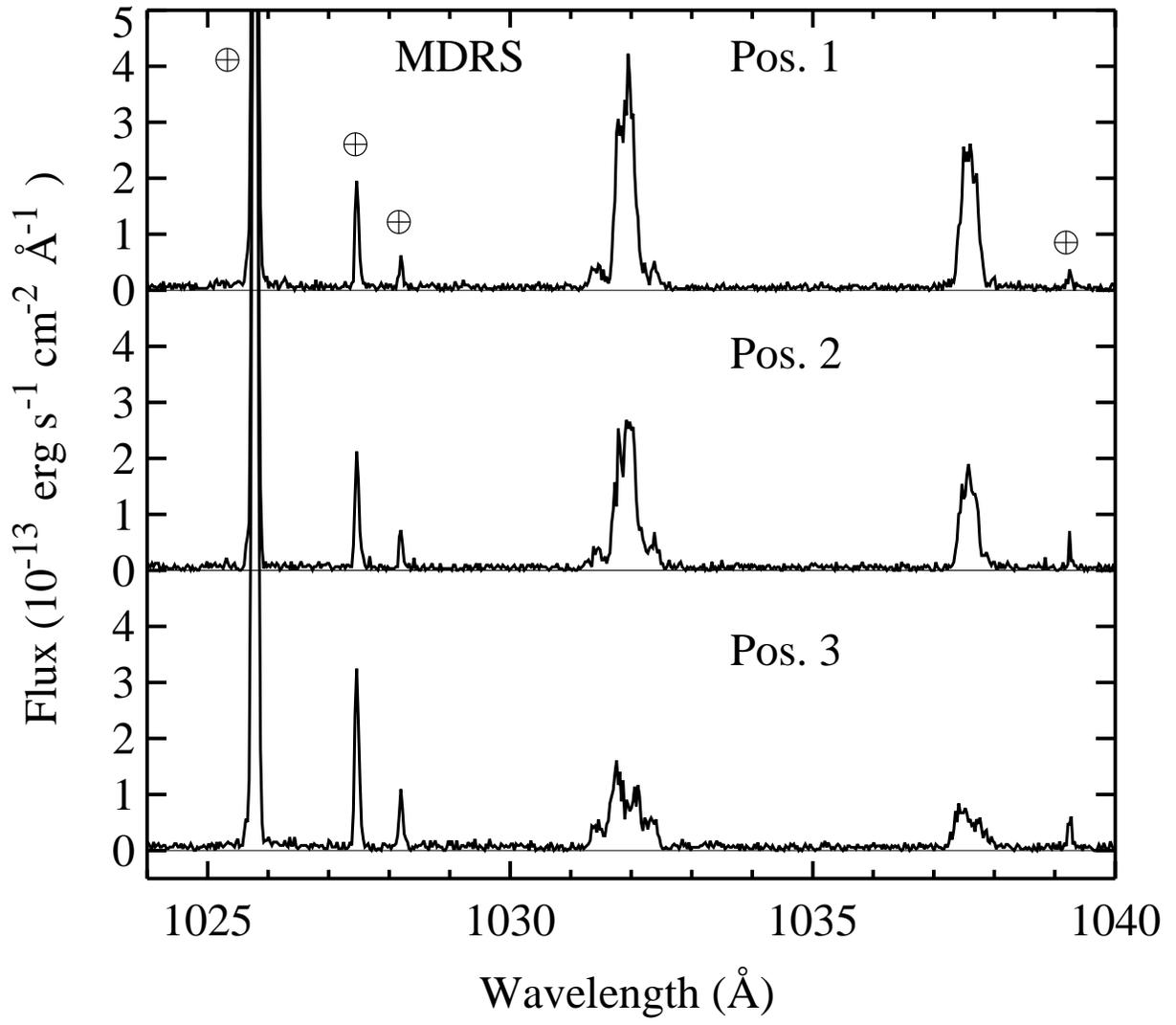
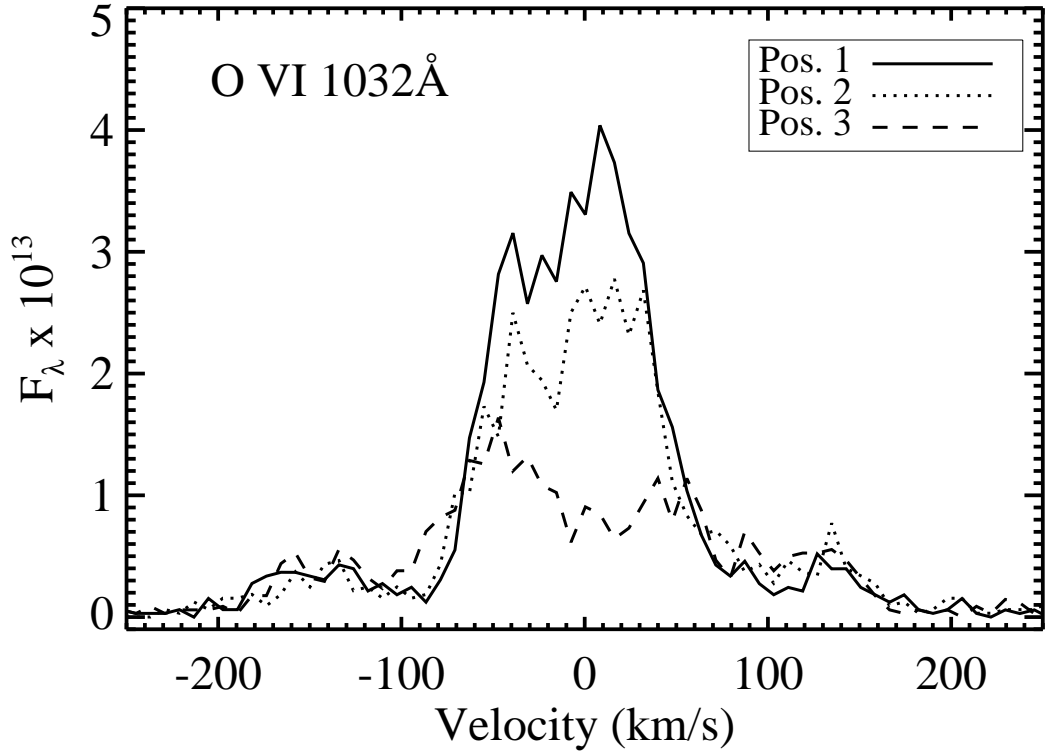
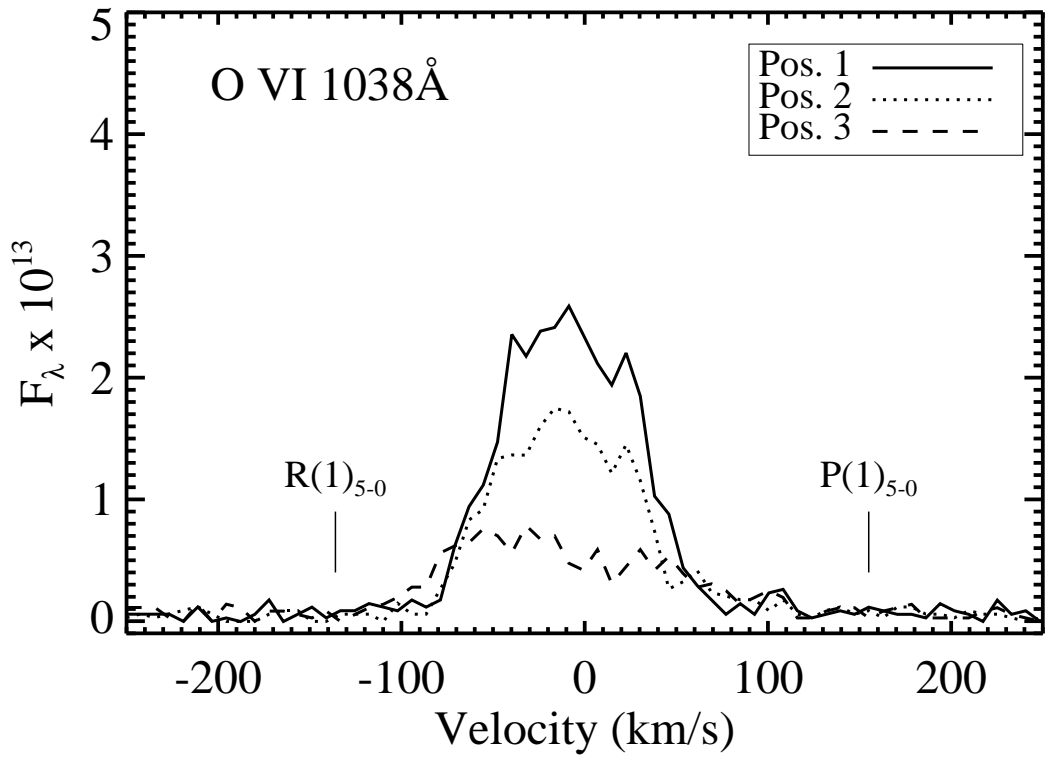
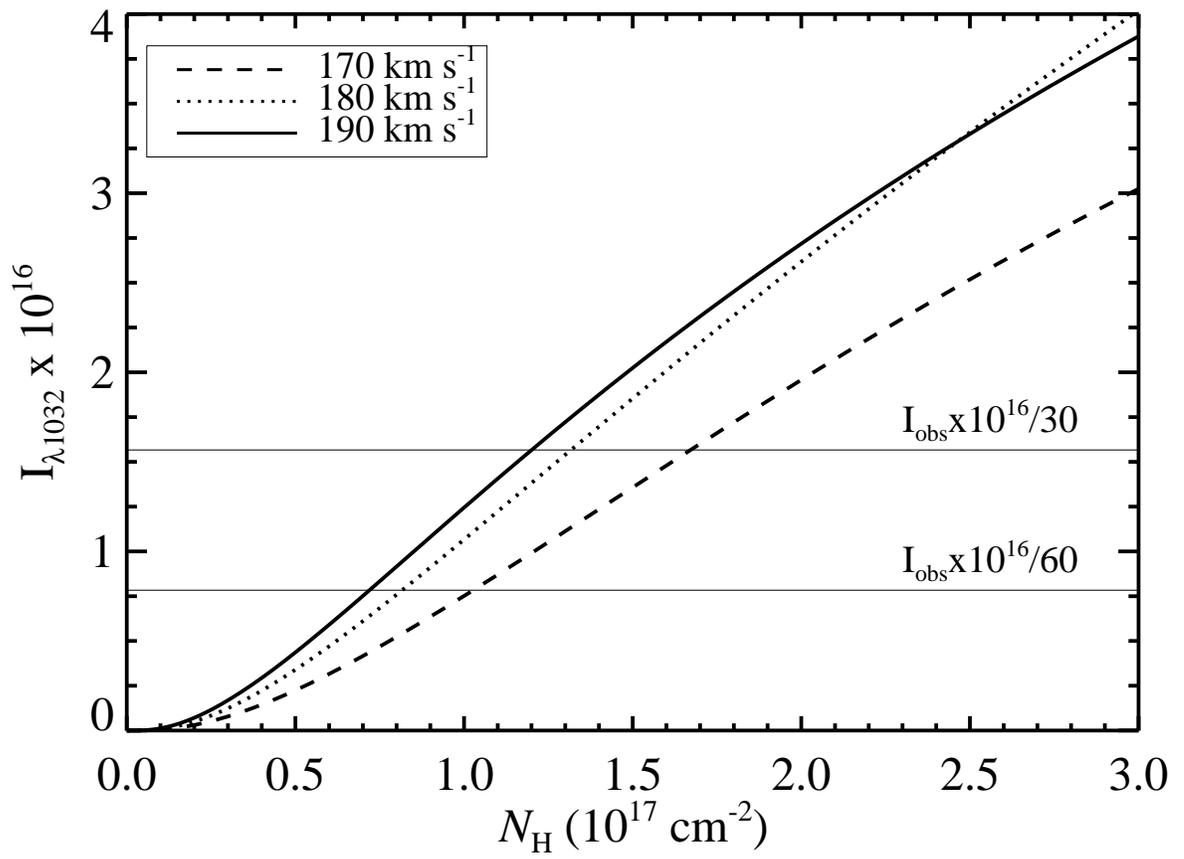
\end{document}